\documentclass[prd,twocolumn,amsmath,amssymb,nofootinbib,floatfix,superscriptaddress]{revtex4}

\usepackage{graphicx,bm,float}
\usepackage[normalem]{ulem} 

\makeatletter
\def\graphicscale{\twocolumn@sw{0.3}{0.4}}
\def\graphicthreescale{\twocolumn@sw{0.3}{0.4}}

\begin{document}

\title{Asymptotic low-temperature behavior of two-dimensional 
RP$^{N-1}$ models}

\author{Claudio Bonati} 
\affiliation{Dipartimento di Fisica dell'Universit\`a di Pisa 
       and INFN, Largo Pontecorvo 3, Pisa, Italy}

\author{Alessio Franchi} 
\affiliation{Dipartimento di Fisica dell'Universit\`a di Pisa,
       Largo Pontecorvo 3, Pisa, Italy}

\author{Andrea Pelissetto}
\affiliation{Dipartimento di Fisica dell'Universit\`a di Roma Sapienza
        and INFN Sezione di Roma I, Roma, Italy}

\author{Ettore Vicari} 
\affiliation{Dipartimento di Fisica dell'Universit\`a di Pisa
       and INFN, Largo Pontecorvo 3, Pisa, Italy}

\date{\today}

\begin{abstract}
We investigate the low-temperature behavior of two-dimensional (2D) RP$^{N-1}$
models, characterized by a global O($N$) symmetry and a local ${\mathbb Z}_2$
symmetry. For $N=3$ we perform large-scale simulations of four different 2D
lattice models: two standard lattice models and two different constrained
models. We also consider a constrained mixed O(3)-RP$^2$ model for values of
the parameters such that vector correlations are always disordered. We find
that all these models show the same finite-size scaling (FSS) behavior, and
therefore belong to the same universality class. However, these FSS curves
differ from those computed in the 2D O(3) $\sigma$ model, suggesting the
existence of a distinct 2D RP$^2$ universality class. We also performed
simulations for $N=4$, and the corresponding FSS results also support the
existence of an RP$^3$ universality class, different from the O(4) one.
\end{abstract}

\maketitle


\section{Introduction}
\label{intro}

Global and local gauge symmetries play a crucial role in the construction of
quantum and statistical field theories, relevant for fundamental
interactions~\cite{Weinberg-book} and emerging phenomena in condensed matter
physics~\cite{Sachdev-19}.  They determine the main features of the model, such
as the phase diagram and the nature of their thermal and quantum phase
transitions.  The critical behavior arising from the interplay between global
and local gauge symmetries has been investigated in several physical contexts.
Paradigmatic examples are the finite-temperature transitions in quantum
chromodynamics, the theory of strong interactions~\cite{PW-84,BPV-03,PV-13},
and in the multicomponent Abelian-Higgs model~\cite{ZJ-book,PV-19}.  In the
case of nonabelian gauge symmetries, the nature of the phase transitions is
mostly determined by the global symmetries, in both three-dimensional (3D) and
two-dimensional (2D) models, while the modes associated with the local gauge
symmetries play only a marginal role, see, e.g.,
Refs.~\cite{PW-84,PV-19,BPV-19-3d,BPV-20-on,BPV-20-2d}.  This is not the case
for the abelian U(1) gauge theories, in which some features of the gauge
group---in particular, the topology of the gauge-field configurations---play an
important role. There is now a wide consensus that, in three dimensions, the
critical behavior depends on the presence/absence of topological defects like
monopoles and hedgehogs and on the compact/noncompact nature of the gauge
fields
\cite{Polyakov-75,SVBSF-04,MV-04,SBSVF-04,BMK-13,NCSOS-15,PV-20-largeN,PV-20-mfcpn}.
Also in the case of antiferromagnetic models, gauge fields apparently play an
important role, and indeed, effective models in which they are integrated out
do not describe their critical behavior \cite{DPV-15}.

RP$^{N-1}$ models represent a class of systems characterized by the
simultaneous presence of a global and a local gauge symmetry. They are
$N$-component vector models that are invariant under global O($N$) and local
${\mathbb Z}_2$ transformations, and they are expected to describe the
universal features of the isotropic-nematic transition in liquid
crystals~\cite{deGennes-book}.  Ferromagnetic RP$^{N-1}$ models in three
dimensions are not particularly interesting as the finite-temperature
transition is of first order~\cite{deGennes-book}.  Critical transition are
instead observed in 3D antiferromagnetic models
~\cite{FMSTV-05,ACFJMRT-05,PTV-18}, whose nature, however has not yet been
fully clarified for $N\ge 4$. Antiferromagnetic models are also relevant 
(but after an analytic continuation to $N=-1$) in the analysis of the 
behavior of spanning forests (see Ref.~\cite{CSS-17} and references 
therein).

In this work we will study the critical behavior of ferromagnetic 2D RP$^{N-1}$
models. Their behavior has been for long controversial, and at present, it is
not yet fully understood, see, e.g.,
Refs.~\cite{CEPS-93,CPS-94,Hasenbusch-96,NWS-96,CHHR-98,DDL-20}.  For $N\ge 3$,
these models are not expected to  undergo finite-temperature 
continuous transitions related to the breaking of the O($N$) symmetry, 
because of the Mermin-Wagner theorem~\cite{MW-66}. A priori, transitions
with quasi-long-range order are possible, but they can be excluded 
using simple comparison arguments \cite{Hasenbusch-96}.
For finite values of the temperature
only first-order transitions are generically allowed, and indeed such
transitions are expected for large values of $N$ \cite{MR-87,SS-01,TS-02}.
Magnetic modes can become critical only in the zero-temperature limit, and in
this limit magnetic correlations increase exponentially, similarly to what
occurs in 2D O($N$) $\sigma$ models.  The nature of such asymptotic
low-temperature behavior has been long debated.
Refs.~\cite{Hasenbusch-96,NWS-96,CHHR-98} reported arguments to support the
claim that RP$^{N-1}$ and O($N$) models belong to the same universality class,
implying the irrelevance of the ${\mathbb Z}_2$ gauge symmetry in the 
zero-temperature limit.  However, these
arguments were never supported by numerical data: in all cases
\cite{CEPS-93,Hasenbusch-96} RP$^{N-1}$ results were in large disagreement with
the predictions obtained by assuming the equivalence of the two classes of
models. A rigorous argument in favor of the equivalence was put forward in
Ref.~\cite{Hasenbusch-96}.  However, it was based on models effectively
designed to eliminate the topological defects, whose presence is essential
to obtain a  different low-temperature behavior for RP$^{N-1}$ models and
O($N$) models.

In this paper we return to the issue of the nature of the
low-temperature critical behavior of 2D RP$^{N-1}$ models. Indeed,
topological defects, even if exponentially rare, can change the
asymptotic nonperturbative behavior of the model (this is the case of
the compact U(1) gauge theory in three dimensions, see
Ref.~\cite{Polyakov-75}). For this purpose we study the finite-size
scaling (FSS) behavior of several different RP$^{N-1}$ models, with
$N=3$ and 4.  If all these models are in the same O($N$) universality
class, we would expect them to have the same FSS behavior as the
standard O($N$) model. If discrepancies are present, in this scenario
they would be interpreted as scaling corrections that would be
therefore nonuniversal, that is they would depend on the model.
Therefore, the results corresponding to the different models should
either fall on top of the O($N$) FSS curves or should all be
different.  As we shall see, this does not occur. The RP$^{N-1}$ data
show universality: all data fall on the same FSS curve, with tiny
differences that would be naturally interpreted as scaling
corrections. The resulting FSS curve is distinctly different from the
corresponding one obtained in the O($N$) vector model. Therefore, the
observed universal behavior supports the existence of an RP$^{N-1}$
distinct universality class.
 
The paper is organized as follows. In Sec.~\ref{models} we report the
Hamiltonians of the two standard RP$^{N-1}$ models we consider. In
Sec.~\ref{sec3} we review the different scenarios for the behavior of
RP$^{N-1}$ models. In Sec.~\ref{fsssec} we define our observables and review
the FSS methods that are used in the numerical analysis of the data. In
Sec.~\ref{rp2res} we present the FSS analyses of the numerical data for the two
models introduced in Sec.~\ref{models}. In Sec.~\ref{sec5} we consider a class
of models introduced in Refs.~\cite{Hasenbusch-96,SP-93}, discuss the rigorous
arguments of Ref.~\cite{Hasenbusch-96}, and present numerical results for this
class of models. Our conclusions are reported in Sec.~\ref{sec6}.

\section{Two-dimensional RP$^{N-1}$ models} \label{models}

RP$^{N-1}$ models are $N$-vector models characterized by a global
O($N$) symmetry and a local ${\mathbb Z}_2$ gauge symmetry.  A lattice
formulation of the RP$^{N-1}$ model on a square lattice can be
obtained by considering real $N$-dimensional vectors ${\bm
  S}_{\bm x}$ of unit length defined on the sites of the lattice (they satisfy ${\bm
  S}_{\bm x} \cdot {\bm S}_{\bm x}=1$) and the Hamiltonian
\begin{eqnarray}
H = - J \sum_{{\bm x}, \mu} ({\bm{S}}_{\bm x} 
\cdot {\bm S}_{{\bm x}+\hat\mu})^2.
\label{hstandard}
\end{eqnarray}
Here $\hat\mu=\hat{1},\hat{2}$ are unit vectors along the lattice
directions and the sum runs over all lattice links.  
The partition function of the system reads
\begin{equation}
Z = \int [d{\bm S}_{\bm x}] e^{-\beta H},\qquad 
\beta\equiv 1/T\,.
\label{partfucrp2}
\end{equation}
Alternatively we may consider a lattice model with an explicit
${\mathbb Z}_2$ gauge variable $\sigma_{{\bm x},\mu}=\pm 1$ associated
with each link.  The Hamiltonian is in this case
\begin{eqnarray}
H_\sigma = - J \sum_{{\bm x}, \mu} 
{\bm{S}}_{\bm x} 
\cdot \sigma_{{\bm x},\mu}  {\bm S}_{{\bm x}+\hat\mu}
\label{hgauge}
\end{eqnarray}
and the partition function reads 
\begin{eqnarray}
&&Z = \int [d {\bm S}_{\bm x}] \sum_{\{\sigma_{{\bm x},\mu}\}} 
e^{-\beta H_\sigma}.
\label{partfunw}
\end{eqnarray}
The fields $\sigma_{{\bm x},\mu}=\pm 1$ can be trivially integrated out,
obtaining the effective model with partition function
\begin{eqnarray}
&& Z = \int [d {\bm S}_{\bm x}] e^{-\beta H_{\sigma,{\rm eff}}}\,, \\
&&H_{\sigma,{\rm eff}} = - \beta^{-1} \sum_{{\bm x},\mu} 
\ln 2 {\rm cosh}(\beta J |{\bm{S}}_{\bm x} 
\cdot {\bm S}_{{\bm x}+\hat\mu}|)\,.
\nonumber
\end{eqnarray}
For $\beta$ large, the expression of the Hamiltonian 
can be simplified obtaining 
\begin{equation}
H_{\sigma,{\rm eff}} = - J \sum_{{\bm x},\mu} |{\bm{S}}_{\bm x}
\cdot {\bm S}_{{\bm x}+\hat\mu}|,
\label{Heffective}
\end{equation}
with corrections that are exponentially small in $\beta$.
These models are invariant under the global O($N$) rotations of the
$N$-component spin variables and under the local ${\mathbb Z}_2$ gauge
transformations ${\bm S}_{\bm x} \to s_{\bm x} {\bm S}_{\bm x}$
[supplemented by $\sigma_{{\bm x},\mu} \to s_{\bm x} \sigma_{{\bm
      x},\mu} s_{{\bm x} + \hat{\mu}}$ for model (\ref{hgauge})] with
$s_{\bm x}= \pm 1$.  We set $J=1$ for both lattice models without loss
of generality.

Due to the ${\mathbb Z}_2$ gauge symmetry, the critical behavior can
be characterized by studying the correlations of the spin-2
gauge-invariant operator
\begin{equation}
Q_{{\bm x}}^{ab} = S_{\bm x}^a S_{\bm x}^b - \frac{1}{N}\delta^{ab}\,.
  \label{qdef}
\end{equation}
In two dimensions, according to the Mermin-Wagner
theorem~\cite{MW-66}, no finite-temperature transition related to the 
breaking of the O($N$) symmetry can occur. Spins order only in the 
limit $\beta\to\infty$. The asymptotic zero-temperature behavior
can be studied using perturbation theory. It predicts 
the emergence of long-range correlations characterized by 
a length scale that increases exponentially in $\beta$,  as it also 
occurs in 2D O($N$) $\sigma$ models.

\section{Different scenarios for the critical behavior of 
RP$^{N-1}$ models}
\label{sec3}

We wish now to present the different scenarios for the behavior of
RP$^{N-1}$ model that have been proposed in the literature.  One
possibility is that these models undergo a transition at finite
temperature and indeed, a first-order transition is predicted for
large values of $N$ \cite{MR-87,SS-01,TS-02}.  In principle, it is
also possible to have a finite-temperature continuous transition,
where energy-energy correlations display long-range order, while
magnetic modes are noncritical in agreement with the Mermin-Wagner
theorem. Such continuous transitions, whose existence was put forward
in Ref.~\cite{NWS-96}, were observed in a class of modified O$(N)$
models. It was proved rigorously that a finite-temperature first-order
transition line occurs in a class of O($N$) and RP$^{N-1}$ models with
nonlinear Hamiltonians \cite{vES-02}. The endpoint of the transition
line is expected to correspond to a continuous finite-temperature
transition in the Ising universality class: this was verified
numerically in Ref.~\cite{BGH-02} for $N=3$ and in the large-$N$ limit
in Ref.~\cite{CMP-05}.  A similar behavior is expected in mixed
O($N$)-RP$^{N-1}$ models for large values of $N$ \cite{MR-87,CMP-05}.

A priori it is also possible that the system has a continuous magnetic 
transition
without the presence of a magnetized low-temperature phase, as it occurs 
for $N=2$. As discussed in Ref.~\cite{Hasenbusch-96} this possibility 
is unlikely. Consider indeed the model with Hamiltonian (\ref{hgauge}): the
role of the $\sigma$ fields is that of adding additional disorder in the 
system and thus we expect (and verify numerically in the following)
that the magnetic correlation lengths in the RP$^{N-1}$ model and in
the corresponding O($N$) model satisfy the inequality 
$\xi_{RP^{N-1}}(\beta) < \xi_{O(N)}(\beta)$. Since $\xi_{O(N)}(\beta)$ is always
finite for finite $\beta$, we can exclude the presence of 
finite-temperature transition with a diverging magnetic correlation length.

At present, there is no indication of the presence of a (continuous or
first-order) finite-temperature transition for $N=3$
\cite{CEPS-93,CPS-94,CHHR-98}. The only scenario that is consistent
with the data is the one in which no transition occurs for finite
$\beta$: a critical behavior is only observed for $\beta\to\infty$.
In this limit the $\beta$-dependence of the observables can be
computed in perturbation theory. For the Hamiltonian (\ref{hstandard})
the perturbative behavior, however, can only be observed for large
values of $\beta$---therefore, for very large correlation
lengths---since perturbative corrections are very large \cite{CP-95}.
For $N=3$ it is practically impossible to verify the perturbative
asymptotic scaling \cite{CP-95}. For perturbative considerations, it
is much more interesting to consider the gauge Hamiltonian
(\ref{hgauge}). From Eq.~(\ref{Heffective}), it is obvious that any
quantity has exactly the same perturbative expansion in the gauge
RP$^{N-1}$ model and in the usual O($N$) model. For instance, if we
consider the infinite-volume correlation length computed from $Q_{\bm
  x}$, the ratio $\xi_{\infty,{\rm O}(N)}/ \xi_{\infty,{\rm RP}^{N-1}}$ should be
constant apart from nonperturbative corrections that decay
exponentially in $\beta$. Moreover, if O($N$) and RP$^{N-1}$ models
have the same asymptotic universal behavior, the ratio should approach
one as $\beta \to\infty$.

The nonperturbative behavior is a different issue. 
As discussed in Ref.~\cite{CEPS-93,CPS-94}, the question of the
equivalence of RP$^{N-1}$ and O$(N)$ is directly related to the
question of the nature of their lowest-energy excitations. If an
RP$^{N-1}$ universality class exists, one expects the lowest-energy
excitations to be associated with the bilinear field $Q_{\bm x}$.  On
the other hand, if such a universality class does not exist, the
lowest-energy excitation are associated with vector modes as in the
standard O($N$) model. In Refs.~\cite{CEPS-93,CPS-94}, the authors
considered this possibility unlikely, as the vector correlation
function $\langle {\bm S}_x \cdot {\bm S}_y\rangle$ is trivial in
RP$^{N-1}$ models. However, in the context of these models, it is
probably more appropriate to consider the gauge-invariant correlation
function
\begin{equation}
G_P(x,y) = {\bm S}_x\cdot {\bm S}_y \prod_{\ell\in P} \, 
    \hbox{sign } ({\bm S}\cdot {\bm S})_\ell,
\end{equation}
where $P$ is a path connecting $x$ and $y$, $\ell$ is a link belonging
to the path, and $({\bm S}\cdot {\bm S})_\ell$ is the scalar product
of the two spins at the endpoints of the link.  In models in which
there is an explicit gauge field $\sigma_{x,\mu}$, $\hbox{sign}({\bm
  S}\cdot {\bm S})_\ell$ can be replaced by $\sigma_\ell$.  For
continuous gauge groups (for instance, in the case of CP$^{N-1}$
models) this correlation function is not critical, even for $\beta\to
\infty$ \cite{DV-80,BNS-81,Aoyama-82,CR-92}. Indeed, local string
fluctuations always add up to give rise to an exponential decay $e^{-a
  L_P}$, where $L_P$ is the length of the path $P$ and $a$ is a
path-independent constant.  In our case, the gauge group is discrete
and therefore the behavior of strings of $\sigma$ fields and
of the correlation function $G_P(x,y)$ is less clear.

The possible presence of two distinct universality classes, the O($N$)
and the RP$^{N-1}$ universality class, is related with the behavior of
the effective ${\mathbb Z}_2$ excitations associated with the field
$\sigma_{{\bm x},\mu}$.  In models in which there are no explicit
gauge fields, one can equivalently define
\begin{equation}
   \sigma_{{\bm x},\mu} = \hbox{sign}({\bm S}_{\bm x}\cdot {\bm S}_{{\bm x}+ \hat{\mu}}).
\end{equation}
The relevant variable is the plaquette 
\begin{equation}
\Pi_{\bm x} = \sigma_{{\bm x},1} \sigma_{{\bm x},2} \sigma_{{\bm x} + \hat{1},2} 
     \sigma_{{\bm x} + \hat{2},1}.
\end{equation}
If $\Pi_{\bm x} = 1$ for all sites, in infinite volume (in a finite
volume there are some subtleties~\cite{Hasenbusch-96}, see below) we
can write $\sigma_{{\bm x},\mu} = \tau_{\bm x} \tau_{{\bm
    x}+\hat\mu}$, where $\tau_{\bm x}$ is an Ising spin defined on the
sites of the lattice.  In this case, O($N$) and RP$^{N-1}$ models are
equivalent \cite{Hasenbusch-96,NWS-96}.  Thus, the existence of an
RP$^{N-1}$ universality class depends on the density of the plaquettes
with $\Pi_{\bm x} = -1$ (we will call them topological defects).  This
problem has never been addressed quantitatively, although simple
calculations show that the outcome may be action dependent
\cite{NWS-96}. In particular, one can devise RP$^{N-1}$ models
\cite{Hasenbusch-96} (we will come back to this issue in
Sec.~\ref{sec5}) such that one can rigorously prove that defects are
absent in the asymptotic regime in which the system orders. Therefore,
they have the same critical behavior as the usual O($N$) vector model.

Finally, we mention the scenario proposed by Catterall {\em et al.}
\cite{CHHR-98}. They considered a variant of the gauge RP$^{N-1}$ model
obtained by adding a term $\mu \sum_x \Pi_x$ to the action, 
where $\mu$ plays the 
role of a chemical potential for the defects. For this action they
identified numerically 
a specific renormalization-group trajectory which flows to a 
``vorticity" fixed point and apparently attracts the 
renormalization-group trajectories
for the standard RP$^{N-1}$ gauge model. They conjectured 
that this specific trajectory is 
responsible for the observed quasi-universal
behavior \cite{CHHR-98,Hasenbusch-private}, which is expected to hold
only when $\xi$ is less than a crossover correlation length $\xi^{\rm
cross}$. For $\xi \gtrsim \xi^{\rm cross}$, O($N$) behavior should instead be
observed.
Such a scenario might explain the observed phenomenology, but, we think,
it cannot be tested 
numerically since the crossover correlation length is enormous,
$\xi^{\rm cross} \approx 10^9$. 
Indeed, nonperturbative differences between RP$^{N-1}$ and O($N$) 
models can only observed when $\xi/L\lesssim 1$
(see below), so that any investigation of the nonperturbative behavior for 
$\xi\approx \xi^{\rm cross}$ requires huge  systems with $L\approx 10^9$.

If O($N$) and RP$^{N-1}$ models are equivalent, FSS functions in the
two models can be directly related. One should, however, take into
account that different boundary conditions should be considered in the
two cases. As discussed in Ref.~\cite{Hasenbusch-96}, the FSS
functions of gauge-invariant quantities for the RP$^{N-1}$ model with
periodic boundary conditions should be the same as those of the O($N$)
model with periodic/antiperiodic boundary conditions. The same
argument of Ref.~\cite{Hasenbusch-96} can be used to prove the
equivalence of the RP$^{N-1}$ FSS functions with those of the O($N$)
model with link-fluctuating boundary conditions (LFBC). To define it,
consider a cubic lattice with periodic boundary conditions and divide
the set of lattice links into two disjoint subsets ${\cal B}$ and
${\cal C}$.  We indicate with ${\cal B}$ the set of boundary links
connecting points $ {\bm x} = (L,m) $, ${\bm y} = (1,m) $, and points
$ {\bm x} = (m,L) $, ${\bm y} = (m,1) $ ($m=1,\ldots L)$; ${\cal C}$
corresponds to the set of internal links (the lattice links that do
not belong to $\cal B$). The O($N$) model with LFBC is defined by the
Hamiltonian
\begin{equation}
H_{\rm lf} = - 
\sum_{\langle {\bm x}, \mu\rangle \in {\cal B}} 
{\bm{S}}_{\bm x} \cdot \sigma_{{\bm x},\mu}  {\bm S}_{{\bm x}+\hat\mu} - 
\sum_{\langle {\bm x}, \mu\rangle \in {\cal C}} 
{\bm{S}}_{\bm x} \cdot {\bm S}_{{\bm x}+\hat\mu}. 
\label{HLFBC1}
\end{equation}
We can also consider an equivalent  Hamiltonian, which is the analogue 
of Hamiltonian (\ref{hstandard}):
\begin{equation}
H_{\rm lf2} = - 
\sum_{\langle {\bm x}, \mu\rangle \in {\cal B}} 
({\bm{S}}_{\bm x} \cdot {\bm S}_{{\bm x}+\hat\mu})^2  - 
\sum_{\langle {\bm x}, \mu\rangle \in {\cal C}} 
{\bm{S}}_{\bm x} \cdot {\bm S}_{{\bm x}+\hat\mu}.
\label{HLFBC2}
\end{equation}

\section{Finite-size scaling in the zero-temperature limit}
\label{fsssec}

In this paper we investigate the nature of the asymptotic
large-$\beta$ behavior of the lattice RP$^{N-1}$ models.  For this
purpose we consider RP$^{N-1}$ models on a square lattice of linear
size $L$ with periodic boundary conditions.

We mostly focus on correlations of the gauge-invariant local variable
$Q_{\bm x}^{ab}$ defined in Eq.~(\ref{qdef}), which is a symmetric and
traceless matrix.  Its two-point correlation function is defined as
\begin{equation}
G({\bm x}-{\bm y}) = \langle {\rm Tr}\, Q_{\bm x} Q_{\bm y} \rangle\,,  
\label{gxyp}
\end{equation}
where the translation invariance of the system has been taken into
account.  The susceptibility and the correlation length are defined as
$\chi=\sum_{\bm x} G({\bm x})$ and
\begin{eqnarray}
\xi^2 \equiv  \frac{1}{4 \sin^2 (\pi/L)}
\frac{\widetilde{G}({\bm 0}) - \widetilde{G}({\bm p}_m)}{\widetilde{G}({\bm p}_m)}\,,
\label{xidefpb}
\end{eqnarray}
where $\widetilde{G}({\bm p})=\sum_{{\bm x}} e^{i{\bm p}\cdot {\bm x}}
G({\bm x})$ is the Fourier transform of $G({\bm x})$, and ${\bm p}_m =
(2\pi/L,0)$.  We also consider the Binder parameter defined as
\begin{equation}
U = \frac{\langle \mu_2^2\rangle}{ \langle \mu_2 \rangle^2} \,, \qquad
\mu_2 = 
\frac{1}{V^2}\sum_{{\bm x},{\bm y}} {\rm Tr}\,Q_{\bm x} Q_{\bm y}\,,
\label{binderdef}
\end{equation}
where $V=L^2$ is the volume.  To determine the universal features of
the asymptotic zero-temperature behavior we use a FSS approach
~\cite{FB-72,Barber-83,Privman-90,PHA-91,PV-02}.  At
finite-temperature continuous transitions the FSS limit is obtained by
taking $\beta\to \beta_c$ and $L\to\infty$ keeping $X \equiv
(\beta-\beta_c)L^{1/\nu}$ fixed, where $\beta_c$ is the inverse
critical temperature and $\nu$ is the correlation-length exponent.
Any renormalization-group invariant quantity $R$, such as the ratio
\begin{equation}
R_\xi\equiv \xi/L
\label{rxidef}
\end{equation}
and the Binder parameter $U$, is expected to asymptotically behave as
$R(\beta,L) = f_R(X)+O(L^{-\omega})$, where $\omega$ is a universal
exponent.  The scaling function $f_R(X)$ is universal apart from a
trivial normalization of its argument; it only depends on the shape of
the lattice and on the boundary conditions.  Since $R_\xi$ is
generally monotonic, we can also
write~\cite{Barber-83,Privman-90,PHA-91,PV-02},
\begin{equation}
R(\beta,L) = F_R(R_\xi) + O(L^{-\omega}),
\label{r12sca}
\end{equation}
where $F_R$ is a universal scaling function.  Eq.~(\ref{r12sca}) is
particularly convenient, as it allows a direct check of universality,
without the need of tuning any parameter. Moreover, it applies
directly, without any change, to two-dimensional asymptotically free
models~\cite{ZJ-book}, in which a critical behavior is only obtained
in the limit $\beta \to \infty$, see
Refs.~\cite{LWW-91,Kim-93,CEFPS-95,CP-98} and references therein.  In
this case, scaling corrections decay as $L^{-2} \log^p L$, where $p$
cannot be determined in perturbation theory [see Ref.~\cite{CP-98} for
a discussion in the O($N$) model].

In the following, we consider the finite-size behavior of the Binder parameter
$U$ as a function of $R_\xi$: If two models belong to the same universality
class, the Binder parameter $U$ must satisfy the FSS relation (\ref{r12sca})
with the same asymptotic curve $F_U(R_\xi)$.  Universality also implies that
all dimensionless renormalization-group invariant quantities have the same
asymptotic large-$\beta$ behavior, both in the thermodynamic and in the FSS
limit.

\section{Numerical results  for the lattice  RP$^2$ and RP$^3$ models}
\label{rp2res}

\begin{figure}[tbp]
\includegraphics*[scale=\graphicscale]{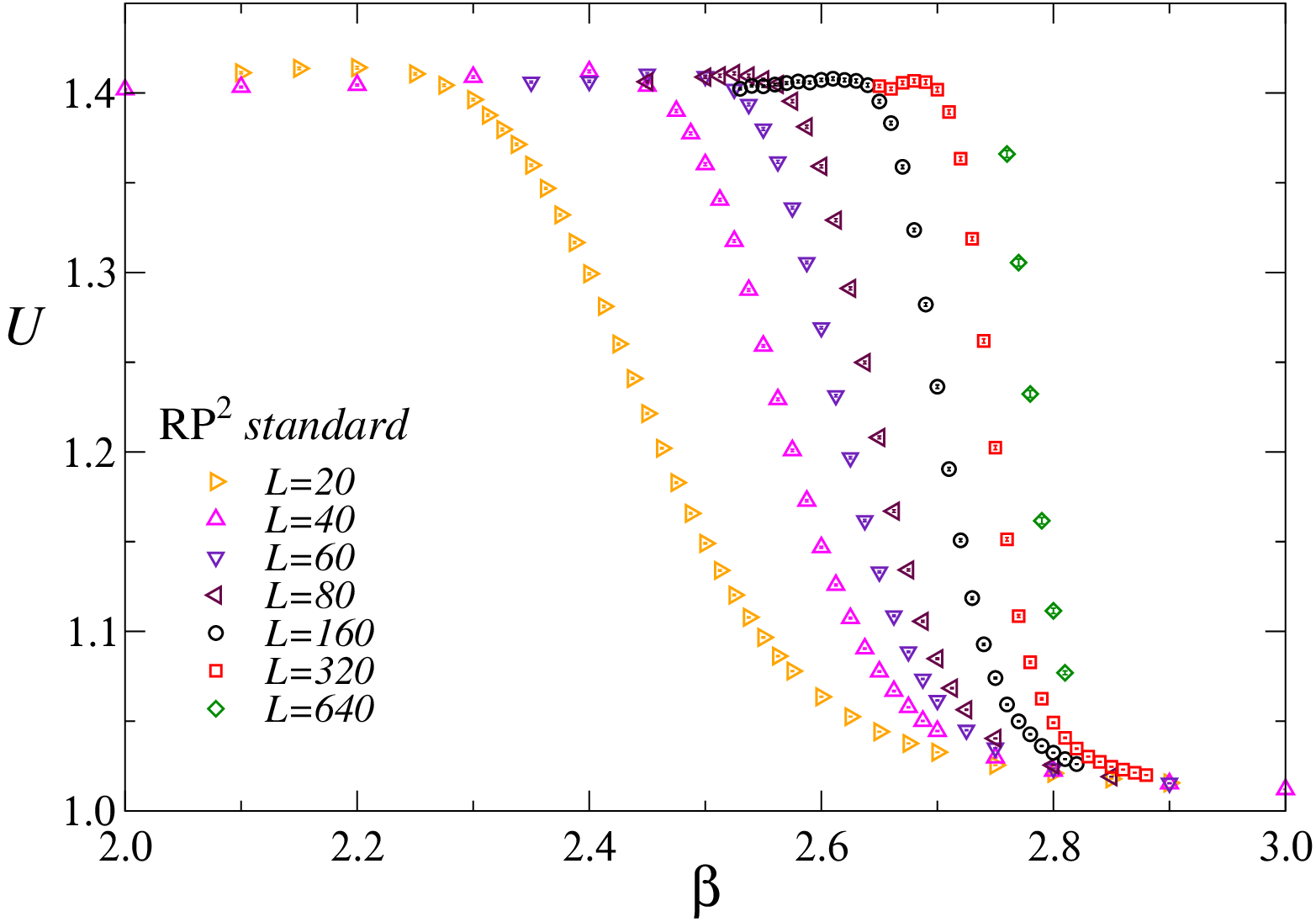}
\includegraphics*[scale=\graphicscale]{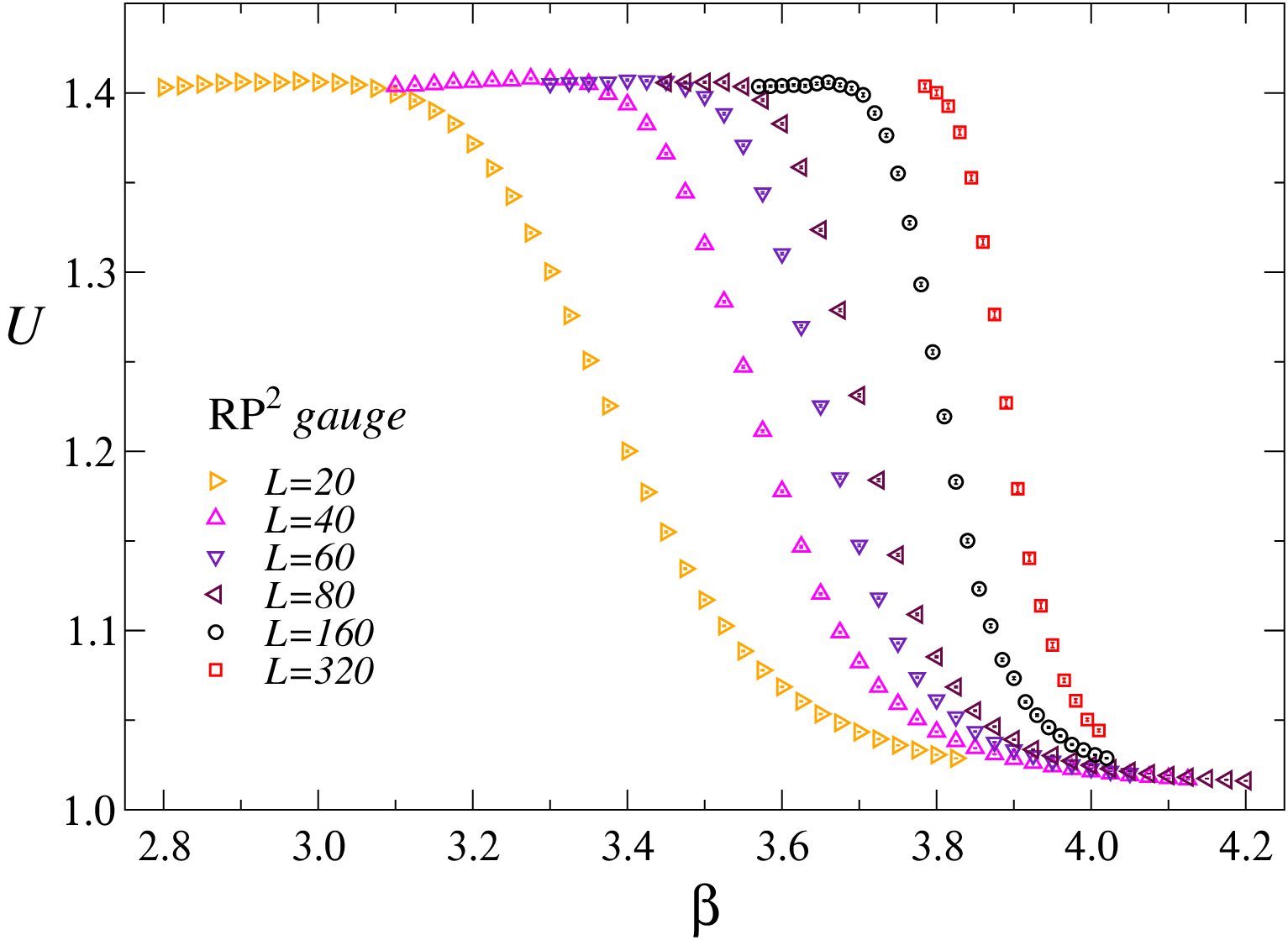}
\caption{ The Binder parameter $U$ vs the inverse temperature $\beta$
  for the standard lattice RP$^2$ model~(\ref{hstandard}) (top panel),
  and for the lattice RP$^2$ model (\ref{hgauge}) with explicit Z$_2$
  gauge link variables (bottom panel).  }
\label{ubeta_rp2}
\end{figure}

\begin{figure}[tbp]
\includegraphics*[scale=\graphicscale]{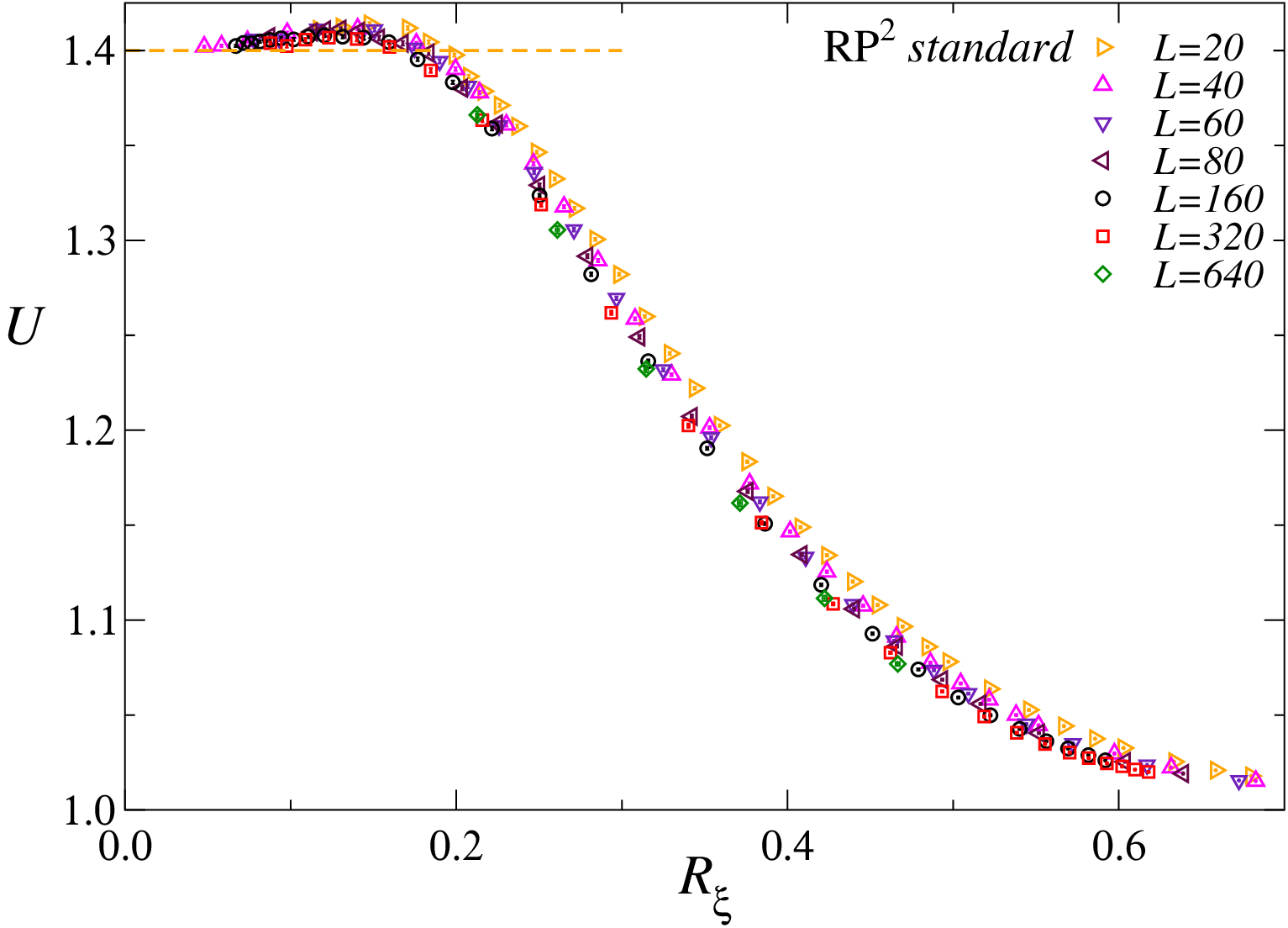}
\includegraphics*[scale=\graphicscale]{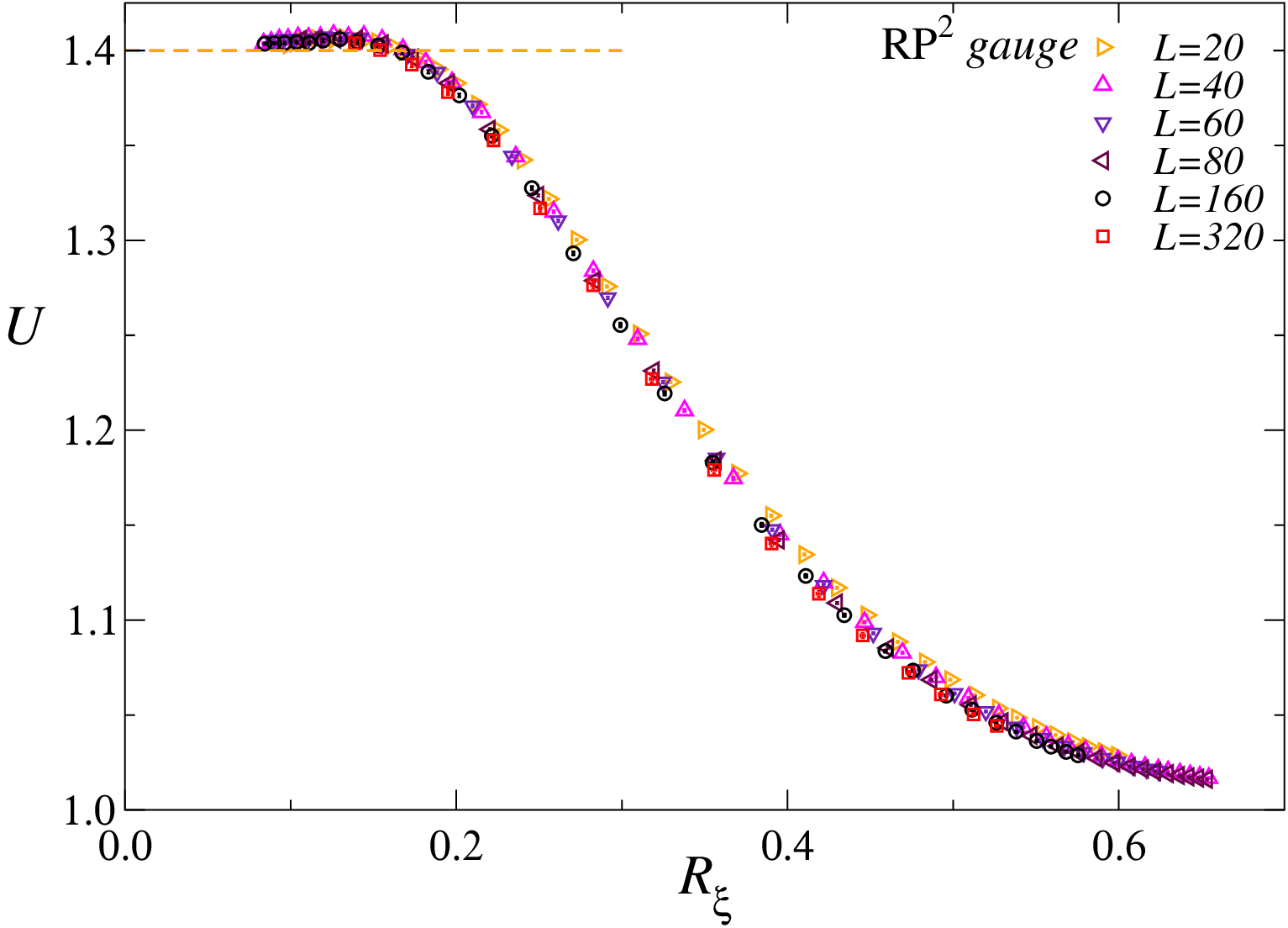}
\includegraphics*[scale=\graphicscale]{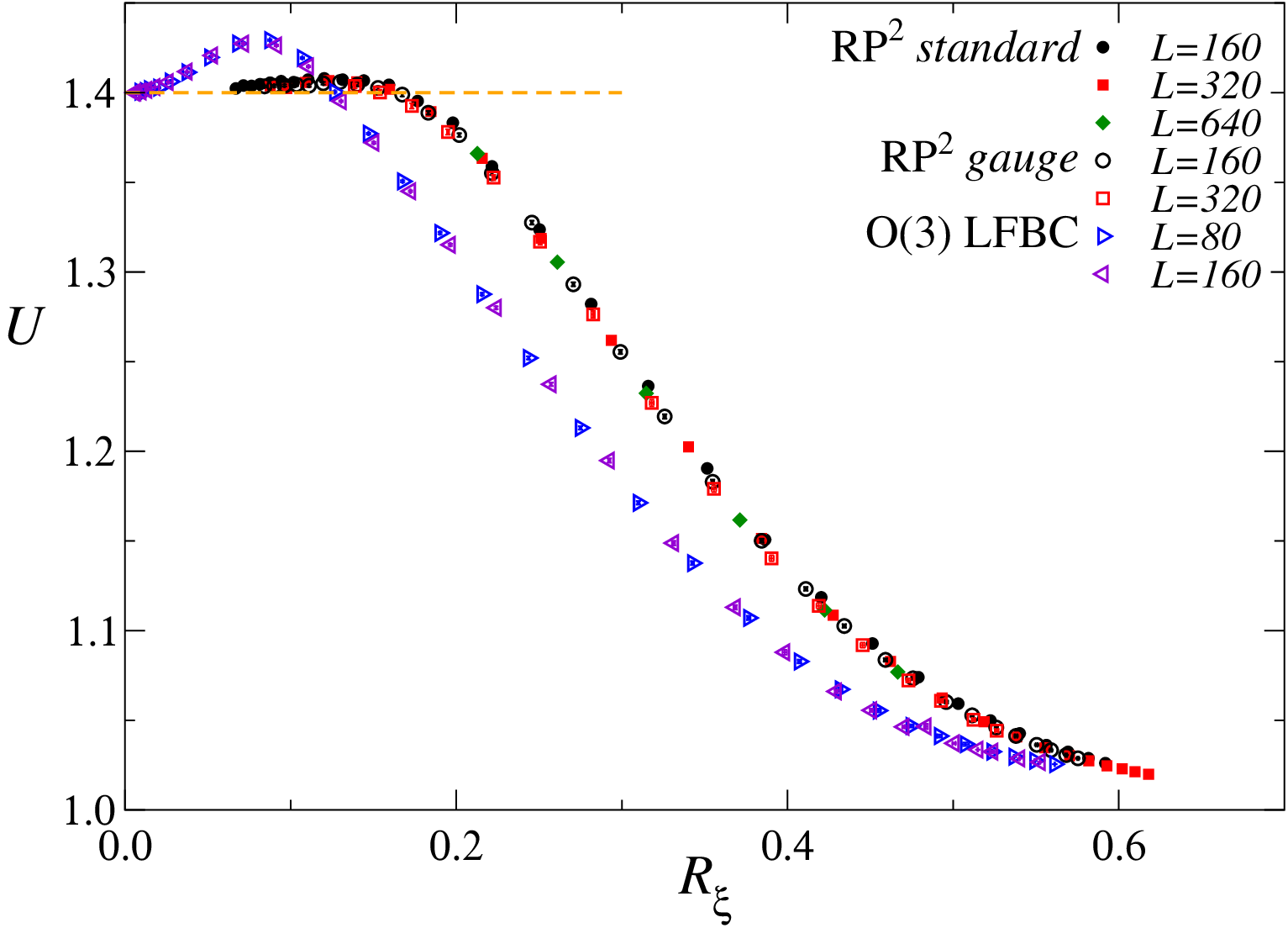}
\caption{Plot of $U$ vs $R_\xi$ for the standard lattice RP$^2$
  model~(\ref{hstandard}) (top panel) and for the lattice RP$^2$ model
  (\ref{hgauge}) with explicit Z$_2$ gauge link variables (middle
  panel). In the lowest panel we report the data of both models (they
  are labelled ``RP$^2$ standard" and ``RP$^2$ gauge") for the largest
  lattices available and the data for the O(3) model with LFBC (``O(3)
  LFBC") with Hamiltonian \ref{HLFBC1}.  The horizontal dashed line
  corresponds to the strong-coupling value $U=7/5$.  }
\label{urxi_rp2}
\end{figure}

To identify the nature of the universal zero-temperature behavior, we have
performed simulations of the lattice RP$^2$ models (\ref{hstandard}) and
(\ref{hgauge}) on a wide range of lattice sizes (up to $L=640$) with periodic
boundary conditions and of the lattice O(3) model (\ref{HLFBC1}) with LFBC (up
to $L=160$). We have also performed a limited study of the case $N=4$,
considering the lattice RP$^3$ model (\ref{hgauge}) and the lattice O(4) model
(\ref{HLFBC1}).  For both the models (\ref{hstandard}) and (\ref{hgauge}) a
standard Metropolis and an overrelaxation algorithm were used to update the
fields ${\bm S}_{\bm x}$, while just Metropolis was used to update
$\sigma_{{\bm x},\mu}$.  For the case of the model (\ref{hstandard}) it is
however numerically convenient to introduce continuous link fields, and rewrite
the Hamiltonian as that of an O($N$) model with annealed gaussian random links
with zero average and variance $\beta/2$, see e.g. \cite{SS-01}.

As explained in Sec.~\ref{fsssec}, we focus on the FSS behaviors of
the Binder parameter $U$ and the ratio $R_\xi\equiv\xi/L$.
Figure~\ref{ubeta_rp2} shows $U$ versus $\beta$ for several lattice
sizes.  The results for the models with Hamiltonians (\ref{hstandard})
and (\ref{hgauge}) are similar.  The Binder parameter $U$ varies
between the strong-coupling value 
\begin{equation}
U = 1 + \frac{4}{(N-1) (N+2)}\,,
\label{ustrong}
\end{equation}
thus $U=7/5$ for $N=3$, and the weak-coupling value $U=1$. We also
note that the datasets corresponding to different lattice sizes do not
show any crossing point, confirming the absence of a
finite-temperature transition.  Analogous results are obtained for the
ratio $R_\xi=\xi/L$.

As already anticipated in Sec.~\ref{fsssec}, our FSS analysis is based
on the determination of the behavior of $U$ as a function of
$R_\xi\equiv \xi/L$. Figure \ref{urxi_rp2} shows the results for
models (\ref{hstandard}) and (\ref{hgauge}). In both cases the data
approach an asymptotic FSS curve as $L$ increases.  Corrections are
small, in particular for the model (\ref{hgauge}).  More
interestingly, the results show a clear evidence of universality: the
data for the two models corresponding to the largest sizes apparently
fall onto the same asymptotic curve, see the lower panel of
Fig.~\ref{urxi_rp2}.  Corrections to the zero-temperature critical
behavior are expected to decay as $L^{-2}$ times a function of $\ln
L$.  In the case of the two lattice RP$^2$ models considered,
convergence is roughly consistent with $L^{-1}$ corrections, likely
because the logarithmic corrections mimic a power term, as often
observed in O($N$) $\sigma$ models, see, e.g., Ref.~\cite{BNW-10} for
a discussion.  Ref.~\cite{DDL-20} suggested that the critical behavior
should be related to that of the O(5) vector model. We have verified
that our curve differs from that computed in the O(5) model. The O(5)
model may turn out to be more appropriate to describe the behavior of
the antiferromagnetic RP$^2$ model, as discussed at length for the
three-dimensional case \cite{FMSTV-05,PTV-18}.

As we mentioned, if O($N$) and RP$^{N-1}$ models belong to the same
universality class, the RP$^{N-1}$ scaling functions for
gauge-invariant quantities should agree with the corresponding ones
for the O($N$) model with LFBC. In the lower panel of
Fig.~\ref{urxi_rp2}, we also report corresponding data [also in the
  O(3) model we consider the correlation length and the Binder
  parameter of the gauge-invariant $Q_{\bm x}$ defined in
  Eq.~(\ref{qdef})] for the model with Hamiltonian (\ref{HLFBC1}).  It
is evident that the O(3) results are very different from those
obtained for the two RP$^2$ models.
This large disagreement, already
noted in Ref.~\cite{Hasenbusch-96} for a different scaling function,
naturally raises some doubts on the scenario in which O($N$) and
RP$^{N-1}$ models have the same nonperturbative
behavior. Note that the differences are only observed for 
$R_\xi \lesssim 0.6$. For larger values of $R_\xi$, 
the O($N$) and RP$^{N-1}$ scaling functions are essentially the same:
we are indeed entering the perturbative regime in which the 
scaling functions can be computed using perturbation theory, which,
as we already mentioned, is expected to be the same for the two classes of 
models. 

As an additional check, we consider the values of the correlation
length as a function of $\beta$. As we mentioned in Sec.~\ref{sec3},
the correlation lengths computed in the standard O(3) model and in the
gauge model at the same value of $\beta$ should be equal, with
corrections that decrease exponentially in $\beta$, if the two models
are asymptotically equivalent. The values of the correlation length in
the O(3) model can be computed using the four-loop results of
Ref.~\cite{CP-95-4loop} (deviations are small \cite{CEFPS-95} and
practically irrelevant for our considerations) and using the estimate
\cite{CEPS-93} $\xi_V/\xi \approx 3.44$, where $\xi_V$ is the
correlation length computed from the vector correlation $\langle {\bm
  S}_{\bm x} \cdot {\bm S}_{\bm y}\rangle$.  For instance, for the
O(3) model we obtain $\xi\approx 1.1\cdot 10^7$ at $\beta = 3.785$ to
be compared with the RP$^2$ result $\xi \approx 44$. Clearly, the
correlation length in the RP$^2$ model is much smaller than what it
should be if the O(3) and the gauge model were nonperturbatively
equivalent.  A similar discrepancy was already noted \cite{CEPS-93-2}
for the standard action (\ref{hstandard}), but its significance was
not clear because of the presence of very large perturbative
corrections decaying as an inverse power of $\beta$ \cite{CP-95}.  Here
instead, perturbative corrections are very small (they are the same as
in the O($N$) model, for which the four-loop expression accurately
reproduces the data for $\beta \approx 2$-3 \cite{CEFPS-95}).
Therefore, the large discrepancy we observe can be hardly interpreted
as a finite-$\beta$ correction.

\begin{figure}[tbp]
\includegraphics*[scale=\graphicscale]{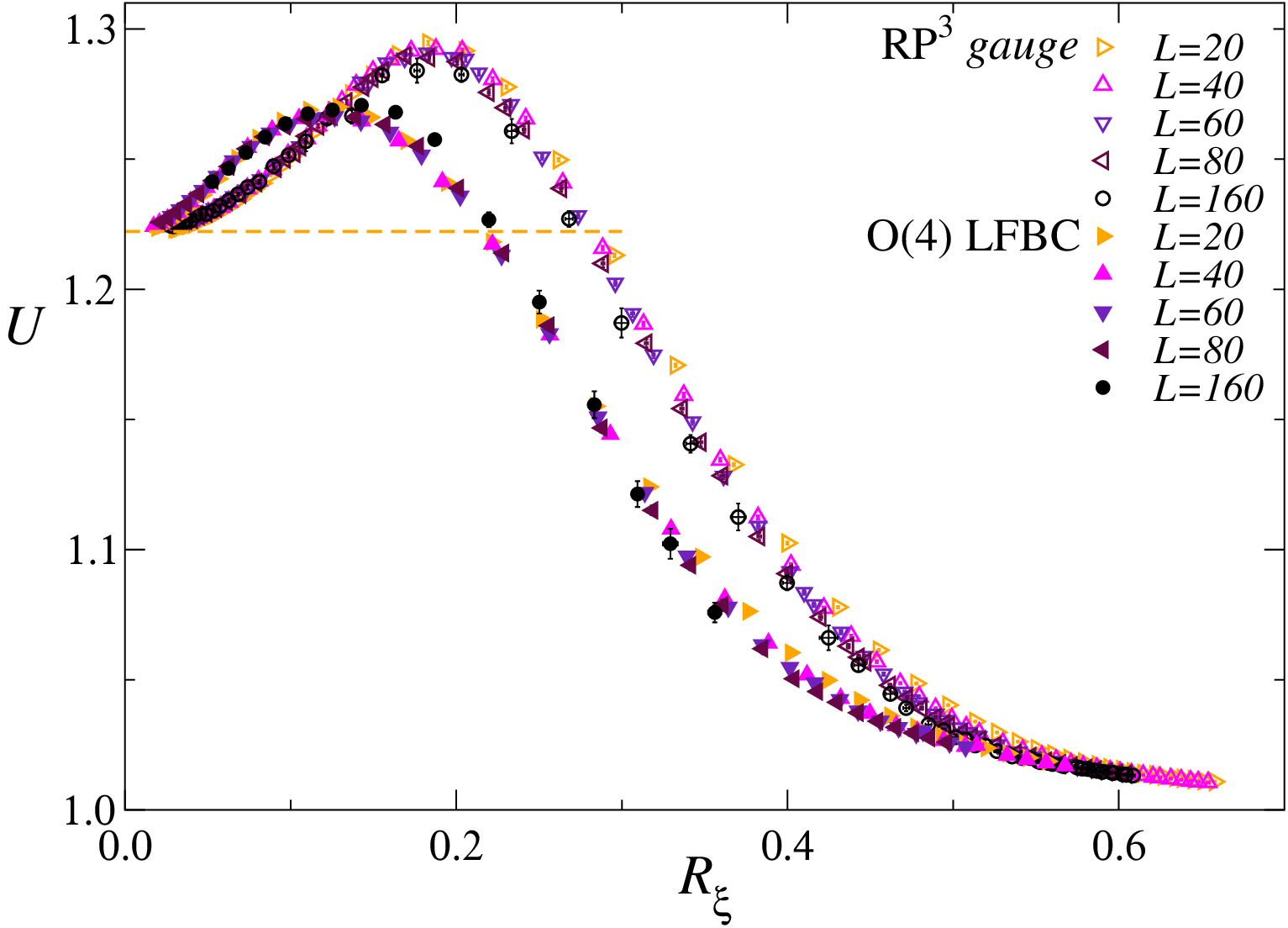}
\caption{Plot of $U$ vs $R_\xi$ for the lattice RP$^3$ model
  (\ref{hgauge}) with explicit Z$_2$ gauge link variables (``RP$^3$ gauge") 
  and for the O(4) model with LFBC (``O(4) LFBC") with Hamiltonian (\ref{HLFBC1}).  
  The horizontal dashed line corresponds to the strong-coupling value $U=11/9$.  }
\label{urxi_rp3}
\end{figure}

To close this section we present some data for case $N=4$: in
Fig.~\ref{urxi_rp3} the FSS of $U$ as a function of $R_{\xi}$ is shown for the
the lattice RP$^3$ model with explicit Z$_2$ gauge link variables
(\ref{hgauge}), and for the O(4) model with link fluctuating boundary
conditions (\ref{HLFBC1}).  These results show that 
also for $N=4$ significant differences
are observed between the RP$^3$ and O(4) data, which point to the existence of a
RP$^3$ fixed point, distinct from that of the O(4) $\sigma$ model.

\section{Patrascioiu-Seiler model} \label{sec5}

In this Section, we discuss another class of models introduced by
Patrascioiu and Seiler \cite{SP-93} and used in the present context by
Hasenbusch \cite{Hasenbusch-96}. We first consider the constrained
RP$^{N-1}$ model, whose partition function is
\begin{equation}
Z = \int [d {\bm S}_{\bm x}]\, 
 \prod_{\langle {\bm x} \mu\rangle} 
   \Theta\left[|{\bm S}_{\bm x} \cdot {\bm S}_{{\bm x} +\hat{\mu}}| - C\right],
\label{RPc}
\end{equation}
where $\Theta(x)$ is the usual Heaviside function, $\Theta(x) = 1,0$
for $x > 0$ and $x < 0$, respectively, and $C$ is a free parameter
that plays the role of $\beta$.  The product extends over all lattice
links. We will also consider a mixed O($N$)-RP$^{N-1}$ model defined
by the partition function
\begin{eqnarray}
 Z &=& \int [d {\bm S}_{\bm x}]\, 
 \prod_{\langle {\bm x} \mu \in {\cal B} \rangle} 
   \Theta\left[|{\bm S}_{\bm x} \cdot {\bm S}_{{\bm x} +\hat\mu}|
- C\right] \times 
\label{RPcmix} \\
 && \quad \prod_{\langle {\bm x} \mu \in {\cal C} \rangle} 
   \left\{ p \Theta\left[{\bm S}_{\bm x} \cdot {\bm S}_{{\bm x} +\hat\mu} - 
      C\right] \right.  \nonumber \\ 
  && \qquad\qquad 
    \left. + 
     (1-p) \Theta\left[|{\bm S}_{\bm x} \cdot {\bm S}_{{\bm x} +\hat\mu}| 
   - C\right] \right\},
\nonumber 
\end{eqnarray}
where $0\le p \le 1$ is a second free parameter, ${\cal B}$ and $\cal
C$ are the sets of boundary and internal links, respectively, as
defined in Sec.~\ref{sec3}.  For $p = 0$, we reobtain model
(\ref{RPc}), while for $p=1$ we obtain an O($N$) model which
corresponds to the standard one with LFBC, see Eq.~(\ref{HLFBC2}).
The parameter $C$ plays the role of temperature.  For $C\to 1$, spins
order, so that this limit corresponds to the limit $\beta\to\infty$ in
the standard case.  However, models with partition functions
(\ref{RPc}) and (\ref{RPcmix}) are not amenable to a perturbative
treatment, so that perturbative considerations on the equivalence of
the different models cannot be used here.  Nonetheless, in the O($N$)
case it has been shown quite precisely that these constrained models
have the same nonperturbative behavior (same continuum limit) as the
standard models \cite{BGPW-10,BNPWW-12}.

In order to have a model in which the geometry of the interactions is
different---so far we have only considered models with
nearest-neighbor interactions---we also consider a Hamiltonian in
which also the spins along the plaquette diagonals interact.  The
partition function is given by
\begin{eqnarray}
Z &=& \int [d{\bm S}_{\bm x}]\, 
 \prod_{{\bm x} \mu} 
  \Theta\left[|{\bm S}_{\bm x} \cdot {\bm S}_{{\bm x} +\hat \mu}| - C\right] 
\times 
\nonumber \\
&& 
 \prod_{{\bm x} d} 
\Theta\left[|{\bm S}_{\bm x} \cdot {\bm S}_{{\bm x} + \hat d}| - C\right],
\label{RPdiag}
\end{eqnarray}
where the vectors $\hat{d}$ are the diagonal vectors $(1,1)$ and $(1,-1)$, 
the first product is over all lattice links and the second one is over 
all lattice plaquette diagonals.

The constrained models are particularly interesting because one can
prove rigorous results concerning their FSS behavior
\cite{Hasenbusch-96}.  For instance, for $C > C^* = \cos\pi/4$, the
behavior of model (\ref{RPcmix}) is independent of $p$.  In
particular, the O($N$) model ($p=1$) with LFBC is equivalent to the
RP$^{N-1}$ model ($p=0$) with periodic boundary conditions. This
implies that the RP$^{N-1}$ model and the O($N$) model have the same
nonperturbative critical behavior. The same is true for model
(\ref{RPdiag}): for $C > \cos\pi/3$, the RP$^{N-1}$ can be exactly
mapped onto an O($N$) model with LFBC. As we shall discuss below, this
exact result should not be taken as a proof that all RP$^{N-1}$ models
are equivalent to O($N$) models.  As the approximate calculations of
Ref.~\cite{NWS-96} show, topological defects may be relevant or
irrelevant depending on the explicit form of the Hamiltonian, and
therefore it is possible that some RP$^{N-1}$ models are not in the
attraction domain of the RP$^{N-1}$ fixed point, if it exists.

As discussed in Ref.~\cite{Hasenbusch-96}, the value $C^*$ corresponds
to very large values of the infinite-volume correlation length. In the
region of sizes in which simulations can be done, $C$ is smaller than
$C^*$: The data that we will show below for model (\ref{RPc}) belong
to the interval $0.5 \le C \lesssim 0.60$, to be compared with $C^*
\approx 0.707$. Thus, for the values of $C$ we consider, the exact
equivalence does not hold.  Therefore, it would not be surprising that
the FSS functions we determine for model (\ref{RPc}) differ somewhat
from the corresponding O($N$) FSS functions. However, if an RP$^{N-1}$
universality class does not exist, we would expect these deviations to
be different from those observed for the more standard RP$^{N-1}$
models discussed in the previous Section.

\begin{figure}[tbp]
\includegraphics*[scale=\graphicscale]{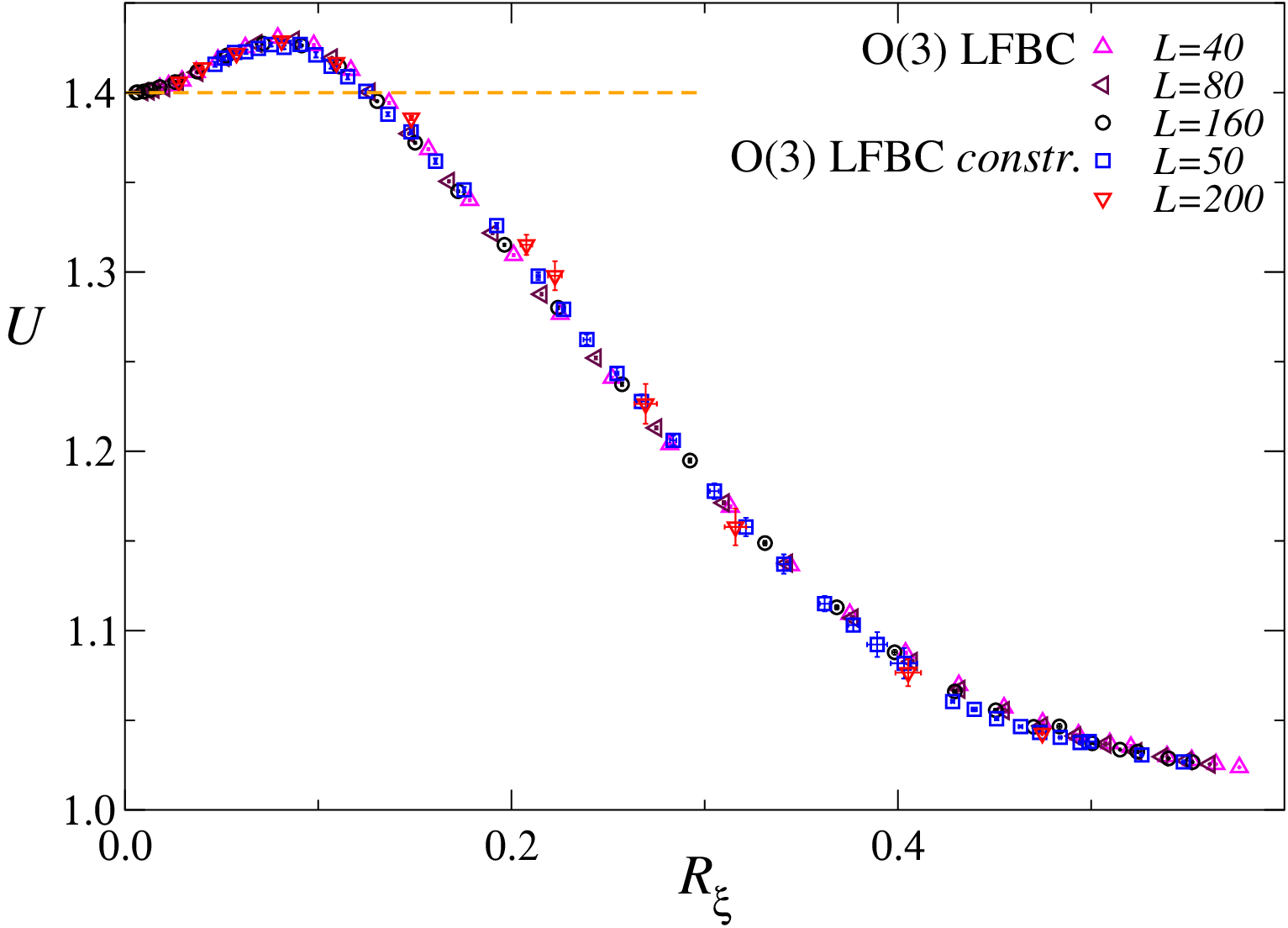}
\caption{ Data of $U$ vs $R_\xi$ for the standard O(3) model with
  fluctuating boundary conditions [Hamiltonian (\ref{HLFBC1})] (``O(3)
  LFBC") and for the model with partition function (\ref{RPcmix}) with
  $p=0$ (``O(3) LFBC constr").  }
\label{urxi_ONLFBC}
\end{figure}

As a first check, we have verified that the constrained O($N$) model
with LFBC (partition function (\ref{RPcmix}) with $p=0$) is equivalent
to the standard O($N$) model with the same boundary conditions
[Hamiltonian (\ref{HLFBC1})]. Results for the Binder parameter versus
$R_\xi$---both quantities are computed using the spin-2 operator
$Q_{\bm x}$, see Eqs.~(\ref{xidefpb}) and (\ref{binderdef})---are
compared in Fig.~\ref{urxi_ONLFBC} (all results presented in this 
section have been obtained using a cluster algorithm \cite{Hasenbusch-96}). 
As expected \cite{BGPW-10}, we
observe very good scaling, indicating that the two models have the
same nonperturbative behavior, although they are not perturbatively
related.

\begin{figure}[tbp]
\includegraphics*[scale=\graphicscale]{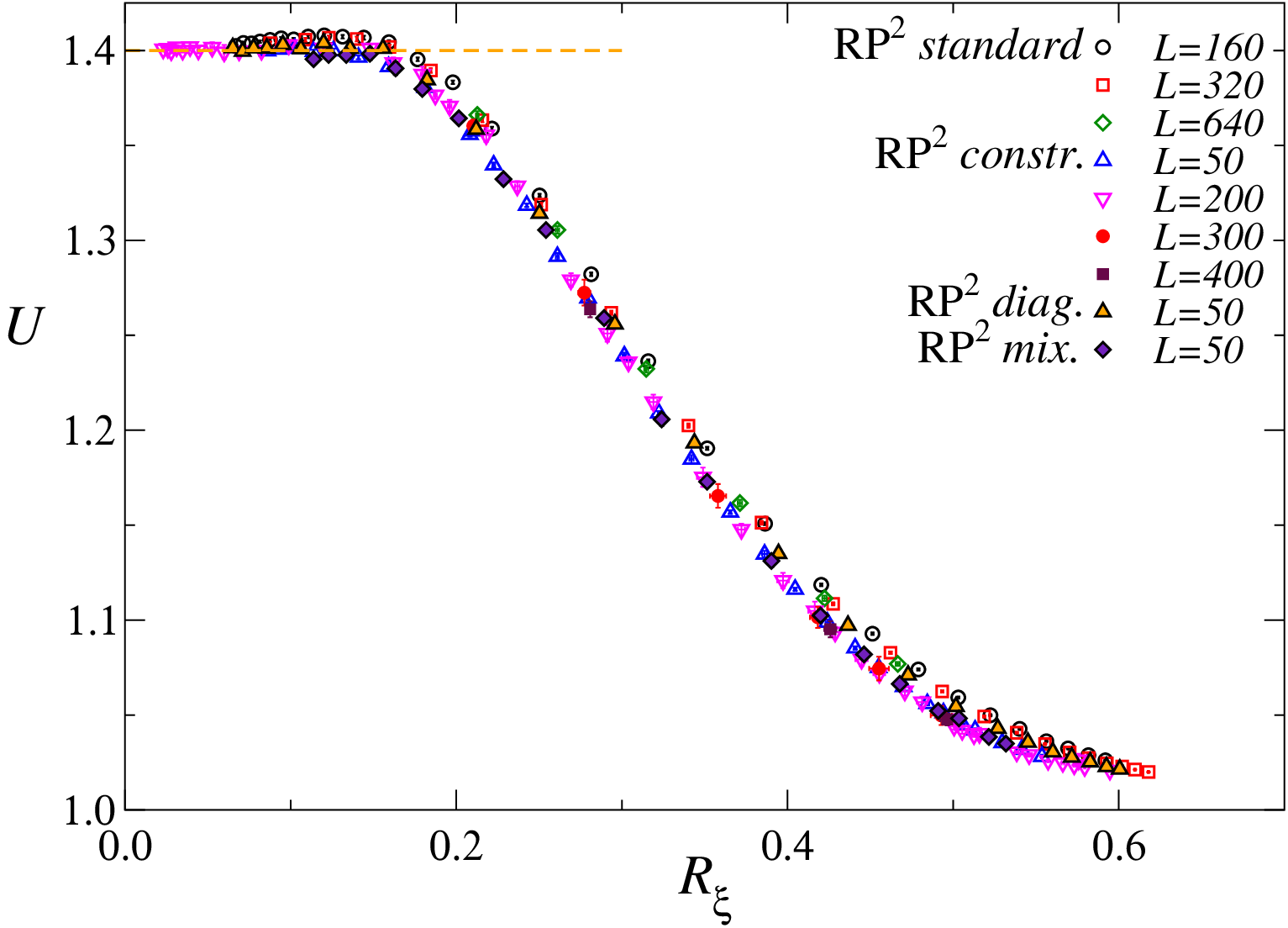}
\caption{ Data of $U$ vs $R_\xi$ for several different RP$^2$ models:
  the model with Hamiltonian~(\ref{hstandard}) (``standard"), the
  model with partition function~(\ref{RPc}) (``constr"), and the model
  with partition function~(\ref{RPdiag}) (``diag").  We also include
  data for the mixed O(3)-RP$^2$ model with partition
  function~(\ref{RPcmix}) and $p=0.2$ (``mix").  }
\label{urxi_rp2c}
\end{figure}

We then turn to the analysis of the behavior of the RP$^{N-1}$ models.
We have performed simulations for the models with partition functions
(\ref{RPc}) (up to $L=400$) and (\ref{RPdiag}) (only $L=50$).  The
estimates of $U$ are plotted vs $R_\xi$ in Fig.~\ref{urxi_rp2c}. The
results for the constrained model (\ref{RPc}) show very good scaling:
all results with $50\le L \le 400$ fall on the same curve within the
statistical errors.  Apparently, corrections to scaling are tiny, a
feature that this model shares with its O($N$) counterpart
\cite{BGPW-10,BNPWW-12}. The results are also compared with those of
the standard RP$^2$ model.  We observe a quite good agreement, that
again would suggest that all these models have the same asymptotic
behavior. Small deviations are observed for $0.3\lesssim R_\xi\lesssim
0.5$, which are of the same order of the deviations observed in the
top panel of Fig.~\ref{urxi_rp2} for the model with Hamiltonian
(\ref{hstandard}).  If an RP$^2$ fixed point exists, they may be
interpreted as scaling corrections.  We also report results for the
model with Hamiltonian (\ref{RPdiag}): the data are again consistent
with the results for the other RP$^2$ models.  Note that the data for
the two models (\ref{RPc}) and (\ref{RPdiag}) correspond to values of
$C$ that are quite different. For $L=50$, the FSS data we show
correspond to $0.50\le C \le 0.60$ in the case of the constrained
model with partition function (\ref{RPc}) and to $0.25 \le C \le 0.35$
for model (\ref{RPdiag}).  Thus, in the two models we consider regions
of configuration space that are quite different.  In spite of that,
the FSS curves are essentially the same.

As an additional check we have considered the mixed O(3)-RP$^2$
model. In Ref.~\cite{CEPS-93,CPS-94} it was suggested that the
RP$^{N-1}$ universal behavior might be observed also in mixed
O($N$)-RP$^{N-1}$. The idea was that of considering the model with
Hamiltonian
\begin{equation}
\beta H = - \beta_V 
  \sum_{x\mu} {\bm S}_{\bm x} \cdot {\bm S}_{{\bm x} + \hat \mu} 
     - \beta_T 
  \sum_{x\mu} ({\bm S}_{\bm x} \cdot {\bm S}_{{\bm x} + \hat \mu})^2,
\end{equation}
and take the limit $\beta_T\to \infty$ at fixed $\beta_V$. In this
limit spins order apart from a sign, i.e., we have $S_{\bm x} =
\hat{n} \tau_x$, where $\tau_x$ is an Ising spin.  In this limit, one
therefore obtains an effective Ising model with $\beta = \beta_V$. It
was therefore conjectured that the limiting theory is different
depending on whether $\beta_V$ is larger or smaller than
$\beta_{c,I}$, the 2D Ising inverse critical temperature. For $\beta_V
< \beta_{c,I}$, the Ising spins are disordered and the behavior is the
same as that of the RP$^{N-1}$ model. In the opposite case, the Ising
spins magnetize and one obtains O($N$) behavior. To verify this
conjecture, we have also performed runs with model
(\ref{RPcmix}). Also in this case, for $C \to 1$, we obtain an
effective Ising model, with inverse temperature $\beta$ related to $p$
by $p = 1 - e^{-2\beta}$.  Thus, for $p < p_c = 1 - e^{-2\beta_{c,I}}
= \sqrt{2} - 1 \approx 0.41$, we expect to observe a behavior
analogous to that observed for the RP$^2$ model, if the conjecture
holds.  Results for $p=0.2$ are reported in Fig.~\ref{urxi_rp2c}.
They scale on top of the RP$^2$ data, apparently confirming the
conjecture.

\begin{figure}[tbp]
\includegraphics*[scale=\graphicscale]{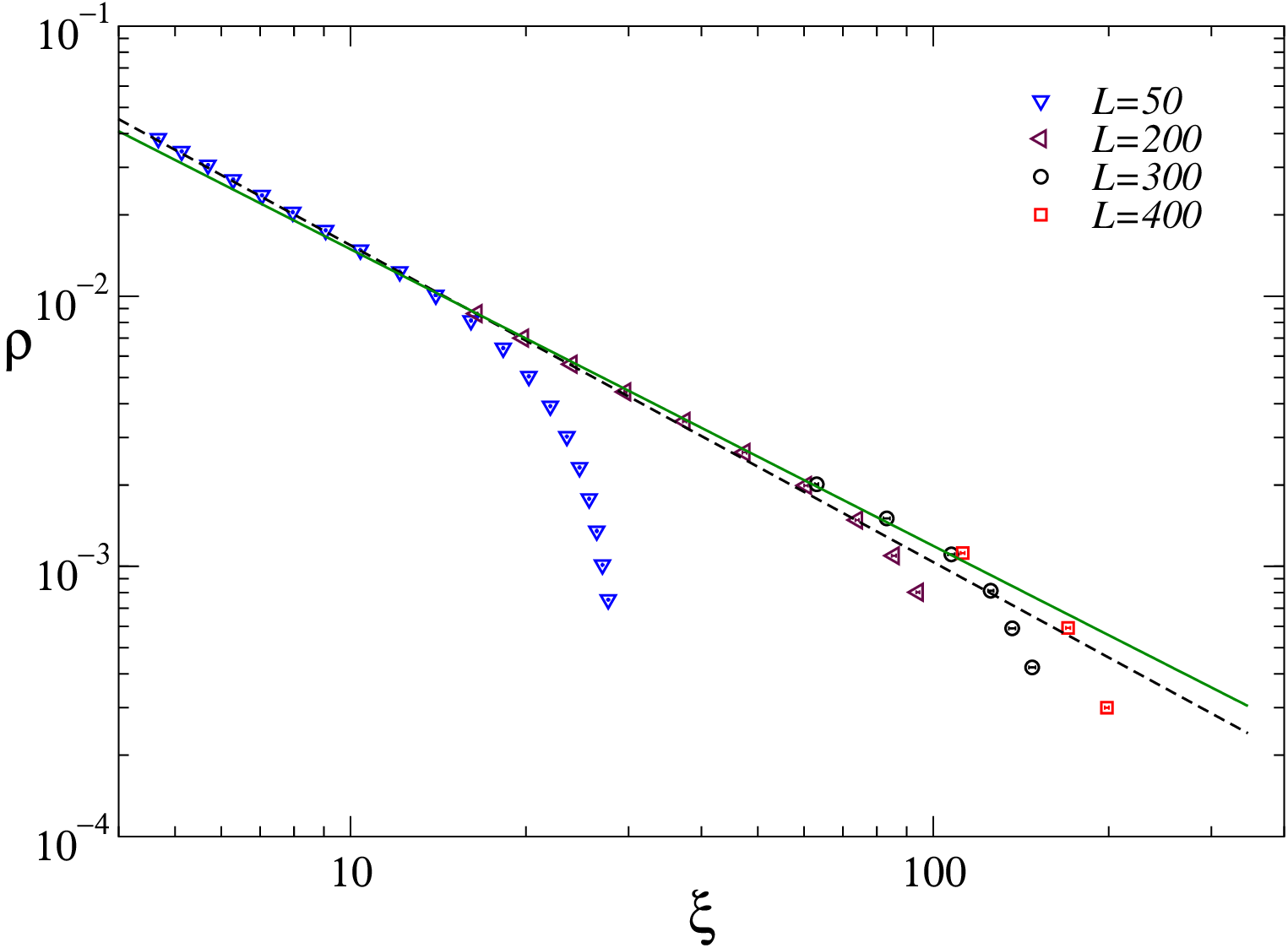}
\caption{Estimates of the defect density $\rho$ versus $\xi$ for the
  model with partition function~(\ref{RPc}).  We also report an
  interpolation of the infinite-volume data with $L=50,200$ (dashed
  line) ($\rho = 0.230 \xi^{-1.173}$) and with $L=200,300$ (continuous
  line) ($\rho = 0.187 \xi^{-1.098}$).  }
\label{densitydefects}
\end{figure}

The results obtained for the constrained models are difficult to
justify if no RP$^{2}$ fixed point exists. Indeed, if the deviations
we observe between the RP$^2$ results and O(3) results are
nonuniversal corrections, we do not see reasons why the RP$^2$ results
are consistently the same, given that we consider models that have
quite different Hamiltonians and interactions. We believe that the
most likely hypothesis is that an RP$^{N-1}$ universality class really
exists. The RP$^{N-1}$ fixed point controls the asymptotic behavior of
models (\ref{hstandard}) and (\ref{hgauge}) and, moreover, it also
controls the apparent scaling behavior we observe in the constrained
models. In the renormalization-group language, for values of $C$ well
below $C^*$, the system is close to the RP$^{N-1}$ fixed point, so
that we observe the RP$^{N-1}$ FSS functions quite precisely. Of
course, as $C$ increases, the RP$^{N-1}$ scaling behavior will
eventually cease to hold and a crossover will eventually occur towards
the asymptotic O($N$) behavior.  But, given that $C^*$ corresponds to
$\xi\sim 10^9$, this will occur when $L$ is much larger than the sizes
we consider.

To provide evidence that the behavior in constrained models for the
current values of $C$ is controlled by the putative RP$^{N-1}$ fixed
point we have analyzed the density of defects
\begin{equation}
\rho = {1\over 2} (1 - \langle \Pi_{\bm x} \rangle ).
\end{equation}
The results are reported in Fig.~\ref{densitydefects} versus the
correlation length $\xi$.  The infinite-volume data ($\xi/L \lesssim
0.2$) scale approximately as a power of $\xi$. A fit of $\rho$ versus
$\xi^{-p}$ gives $p\approx 1$, see Fig.~\ref{densitydefects}. This
result shows that in a correlation volume of size $\xi^2$ the number
of defects increases as $\xi$. Defects are relevant for the values of
$C$ we are considering.

To conclude this section, it is interesting to discuss the phase
structure of the constrained models as a function of $C$. Since the
density of defects is a nontrivial function of $C$ for $C < C^*$ and
vanishes identically for $C>C^*$, the point $C=C^*$ is a
nonanalyticity point of $\rho$.  Given the role that $\rho$ plays in
determining the phase behavior, we expect $C=C^*$ to be a
nonanalyticity point also of the free energy: in other words, $C=C^*$
is a transition point. We do not have informations on the order of
this transition, but the simplest possibility would be that the
transition is of first order. It would separate an approximate
RP$^{N-1}$ phase, where the behavior would be controlled by the nearby
(but unreachable) RP$^{N-1}$ fixed point, from an asymptotic O($N$)
phase. The presence of this transition is a peculiarity of the
constrained models. If we consider Hamiltonians (\ref{hstandard}) and
(\ref{hgauge}), we expect $\rho$ to be nonvanishing for all values of
$\beta$, allowing us to observe the exact asymptotic RP$^{N-1}$
behavior.

\section{Conclusions}\label{sec6}

In this work we analyze the low-temperature behavior of RP$^{N-1}$
models, which are invariant under global O($N$) and local ${\mathbb
  Z}_2$ transformations, with the purpose of understanding whether
these models have a nonperturbative behavior that is different from
that of O($N$) vector models, in spite of the fact that both models
are perturbatively equivalent.  The question effectively boils down to
the question of the relevance/irrelevance of topological ${\mathbb
  Z}_2$ defects. Their density decreases exponentially in $\beta$, but
this does not necessarily imply their irrelevance, as also the
correlation length depends exponentially on the inverse temperature.
The question has been extensively discussed in the '90s, and several
arguments were presented, favoring the existence of a distinct
RP$^{N-1}$ universality class \cite{CEPS-93,CPS-94}, as well as
favoring the equivalence of RP$^{N-1}$ and O($N$) models
\cite{Hasenbusch-96,NWS-96,CHHR-98}.

In recent years there has been a widespread interest in the role that
topology plays in determining the phase behavior of lattice systems.
As an example, we mention here the case of the three-dimensional
Abelian-Higgs model (scalar electrodynamics) and of its limiting case,
the CP$^{N-1}$ model. This model has been extensively studied and
there is now a general consensus that topology plays a crucial role:
The critical behavior depends on the compact/noncompact nature of the
U(1) gauge fields or, equivalently, on the presence/absence of
monopoles
\cite{Polyakov-75,MV-04,SBSVF-04,BMK-13,NCSOS-15,PV-20-largeN,PV-20-mfcpn}.
In particular, the large-$N$ fixed point predicted by the
Abelian-Higgs field theory \cite{HLM-74} can only be observed in
models in which monopoles are suppressed
\cite{PV-20-largeN,PV-20-mfcpn}. With these examples in mind, we have
decided to rivisit the problem, focusing on models with $N=3$. A less
detailed analysis has also been performed for $N=4$.

We present results of large-scale MC simulations of several different
RP$^2$ models. We consider the standard model with Hamiltonian
(\ref{hstandard}) and the one with explicit gauge fields [Hamiltonian
  (\ref{hgauge})] and two models, of the type introduced by
Patrascioiu and Seiler \cite{SP-93}, that we name constrained
models. In such systems there is no perturbative expansion.  However,
in an appropriate limit spins order as in the usual lattice RP$^{N-1}$
models \cite{Hasenbusch-96}.  We consider two variants of the model,
differing by the geometry of the interactions. Finally, we also
consider a mixed O(3)-RP$^2$ model for a value of the parameters such
that it should behave as an RP$^2$ model, according to the discussion
of Ref.~\cite{CEPS-93,CPS-94}. The data obtained from the four
different RP$^2$ models and the mixed O(3)-RP$^2$ model show a
universal FSS behavior.  If we plot the Binder parameter $U$ as a
function of $R_\xi \equiv \xi/L$, all data fall onto a single curve,
with tiny deviations that can be interpreted as scaling
corrections. If the RP$^2$ model has the same nonperturbative behavior
of the O(3) model, the FSS data should fall on top of an appropriate
FSS curve computed in the O(3) model. We have performed this
comparison, observing a large discrepancy, that can be hardly
explained with the presence of nonuniversal size corrections. On the
basis of the numerical data, we thus conclude that an RP$^2$
universality class exists, which is distinct from the O(3)
one. Nonperturbatively, the limiting zero-temperature behaviors for
these two classes of models are therefore different. We have repeated
the analysis for $N=4$, observing analogously large differences. We should 
also remark that additional evidence in favor of two distinct universality 
classes is also provided by the analysis of a class of nonabelian
gauge theories with O($N$) global symmetry that are predicted to have 
the same low-temperature behavior as RP$^{N-1}$ models. The numerical 
data for these gauge models are perfectly consistent with 
the results obtained here for RP$^{N-1}$ models \cite{BFPV-20-gauge}.

The presence of a distinct RP$^{N-1}$ universality class implies that the
topological ${\mathbb Z}_2$ defects are relevant perturbations of the
O(N) fixed point. For $N=3$, we have determined the behavior of the density of
these defects in a particular case, verifying that the number of
defects in a correlation volume apparently increases as $\xi$, which
is consistent with a relevant perturbation.

To conclude, let us mention that our results cannot exclude the scenario
proposed in Ref.~\cite{CHHR-98}, in which the behavior of RP$^{N-1}$ 
models for typical box sizes 
is essentially controlled by a renormalization-group trajectory that 
flows to a vorticity fixed point, which, however,
is never reached. In this scenario, as 
soon as $\xi$ is very large (they estimate 
$\xi \approx 10^9$ for the gauge action) the apparent RP$^{N-1}$  
scaling disappears and O($N$) behavior is obtained. Given the very large 
value of the crossover correlation length, we believe that 
it is in practice impossible 
to distinguish between a $\beta=\infty$ fixed point and a finite-$\beta$ 
fixed point that is relevant up to 
a value of $\beta$ so large that $\xi \approx 10^9$.

\emph{Acknowledgement}. 
We thank Sergio Caracciolo and Martin Hasenbusch for discussions.
Numerical simulations have been performed on the CSN4
cluster of the Scientific Computing Center at INFN-Pisa. Some simulations 
have also been performed on the INFN computer farm in Roma.

\end{document}